# O Método do Operador Estatístico de Não Equilíbrio

Clóves Gonçalves Rodrigues
Polytechnic School, Pontifical Catholic University of Goiás,
74605-010, CP 86, Goiânia, Goiás, Brazil
e-mail: cloves@pucgoias.edu.br

**Abstract**

In this work we describe the Non-Equilibrium Statistical Operator Method (NESOM). The NESOM is a powerful formalism that seems to offer an elegant and concise way for an analytical treatment in the theory of irreversible processes, adequate to deal with a large class of experimental situations, and physically clear picture of irreversible processes. The method invented by D. N. Zubarev is also practical and efficient in the study of the optical and carrier dynamics in semiconductors.
pacs: 67.10.Jn; 05.70.Ln; 68.65.-k; 81.05.Ea

**Resumo**

Neste trabalho, descrevemos o Método do Operador Estatístico de Não Equilíbrio (NESOM). O NESOM é um formalismo poderoso que oferece um caminho elegante, preciso e conciso para um tratamento analítico sobre a teoria dos processos irreversíveis. O NESOM é adequado para lidar com uma grande classe de situações experimentais, fornecendo um imagem fisicamente clara de processos irreversíveis que acontecem no sistema. Este método foi inventado por D. N. Zubarev, e é também bem prático e eficiente no estudo de propriedades ópticas e dinâmicas de portadoras de carga e calor em semicondutores.
Palavras-chave: fenômenos de não equilíbrio; teoria cinética; processos de transporte; processos irreversíveis.

Para a formulação de uma estatística de não equilíbrio que forneça subsídios teóricos à termodinâmica dos processos irreversíveis, isto é, àqueles que se desenvolvem em sistemas não equilibrados, foram propostos um grande número de tratamentos sem que, até o momento, uma formulação geral tenha sido alcançada [1,2]. Existem dois problemas básicos: o primeiro é como conciliar a irreversibilidade do comportamento macroscópico com as leis microscópicas da mecânica, que, como se sabe, são reversíveis e o segundo é como obter equações de evolução e transporte generalizadas que descrevam o comportamento do sistema considerado. Essas equações que relacionam a variação do comportamento temporal médio de grandezas físicas observáveis com o movimento das partículas que o constituem, são equações integro-diferenciais não lineares com amplas aplicações, desde sistemas físico-químicos a biológicos, e sua obtenção permanece um campo de estudo ainda aberto.

A função da mecânica estatística de sistemas fora do equilíbrio é determinar as propriedades termodinâmicas e a evolução temporal dos observáveis macroscópicos de tais sistemas a partir de leis dinâmicas que governam o movimento das partículas constituintes. Seus objetivos básicos são: obter equações de transporte e compreender sua estrutura, entender como se desenvolve a aproximação ao equilíbrio, estudar as propriedades de estados estacionários, calcular os valores instantâneos e a evolução no tempo de quantidades físicas que especifiquem o estado macroscópico do sistema. Vejamos como tratar esses problemas seguindo o formalismo dos ensembles estatísticos para sistemas fora (perto ou longe) do equilíbrio.

Dado um particular sistema físico afastado do equilíbrio estamos interessados no comportamento temporal e espacial de certas quantidades $\{Q_1(\vec{r},t), Q_2(\vec{r},t), \cdots, Q_n(\vec{r},t)\}$ as quais chamaremos macrovariáves (embora elas nem sempre descrevam quantidades macroscópicas no sentido usual). Estas macrovariáves podem ser, por exemplo, número de partículas, energia, magnetização, fluxos de partículas, fluxos de energia, etc. O objetivo central é obter as equações de evolução para estas quantidades. Este conjunto de macrovariáveis tem associado um conjunto de operadores Hermitianos $\{\hat{P}_1(\vec{r},t), \hat{P}_2(\vec{r},t), \cdots, \hat{P}_n(\vec{r},t)\}$ que, na representação de Schroedinger, não dependem do tempo, mas podem depender da posição, isto é, são as densidades locais das grandezas $\hat{P}_j$. Estes operadores podem ser hamiltonianos, operadores densidade espacial de partículas, operadores número de partículas, etc. Chamaremos estas quantidades de *variáveis dinâmicas de base*. A relação entre as macrovariáveis e as variáveis dinâmicas de base está dada na forma usual por

$$Q_j(\vec{r},t) = \text{Tr}\{\hat{P}_j(\vec{r})\hat{\rho}(t)\}, \qquad (1)$$

sendo $j = 1, 2, \ldots, n$, onde $\hat{\rho}(t)$ designa o *Operador Estatístico de Não equilíbrio* definido sobre o espaço de Hilbert dos estados quânticos do sistema e $\text{Tr}\{\hat{R}\}$ é o traço do operador $\hat{R}$. No limite clássico, $\hat{\rho}(t)$ é uma função real definida no espaço de fase $\Gamma$ do sistema e $\text{Tr}\{\hat{R}\}$ deve ser interpretado como uma integral da função $\hat{R}$ sobre $\Gamma$).

Um caminho associado ao Método do Operador Estatístico de Não Equilíbrio (NESOM, na acrossemia em inglês), o qual mostra estar proximamente conectado com a termodinâmica fenomenológica irreversível [3], é baseado na separação do Hamiltoniano total do sistema em duas partes

$$\hat{H} = \hat{H}_0 + \hat{H}_1 , \qquad (2)$$

onde $\hat{H}_0$ é chamado parte relevante ou secular composta das energia cinéticas das partículas do sistema e a parte das interações suficientemente fortes relacionadas com os efeitos de relaxação rápidos [4,5], isto é, com tempos de relaxação muito menores que o tempo característico $\Delta t$ do experimento (usualmente a resolução temporal do instrumento de medida). Por outro lado $\hat{H}_1$ contém o resto do hamiltoniano do sistema, ou seja, as interações que produzem efeitos de relaxação lentos, isto é, com tempos de relaxação suficientemente longos ($> \Delta t$). Este procedimento de separação é fundamental e está baseado no princípio de Bogoliubov sobre os tempos de decaimento de correlações [6], o que permite definir uma hierarquia de tempos de relaxação no sistema dissipativo sob consideração.

Uma das questões fundamentais da teoria é a escolha apropriada das variáveis de base. Evidentemente o conjunto $\{Q_j(\vec{r}, t)\}$ deve conter, em particular, todas as quantidades que estamos interessados em descrever. Não obstante não existe um critério geral que permita obter o conjunto completo de variáveis de base para descrever adequadamente um determinado problema. Existe, porém, um critério auxiliar na escolha destas variáveis; este critério, chamado condição de fechamento consiste em que, dado $\hat{H}_0$, as variáveis microdinâmicas de base, os operadores $\hat{P}_j$ devem satisfazer [5]

$$[\hat{H}_0, \hat{P}_j(\vec{r})] = \sum_{k=1}^{N} \Omega_{jk} \hat{P}_k(\vec{r}) , \qquad (3)$$

com $j = 1, 2, \ldots, N$, onde $\Omega_{kl}$ são em geral números, ou eventualmente funções da posição $\vec{r}$ ou operadores diferenciais. Esta condição estabelece que para uma adequada descrição do sistema devemos considerar todas as variáveis de base que satisfazem as $N$ relações dadas pela equação (3). De alguma forma oferece uma condição de necessidade (mas não de suficiência) para a escolha das macrovariáveis, isto é, a definição do espaço de estados macroscópicos (termodinâmico) do sistema, dadas pela Eq. (1).

Uma vez escolhidas as variáveis de base devemos obter uma expressão para o operador estatístico $\hat{\rho}(t)$. Os detalhes do procedimento para a construção deste operador podem ser encontrados na extensa bibliografia sobre o tema (veja, por exemplo, R. Luzzi, A. R. Vasconcellos, J. G. P. Ramos, *Predictive Statistical Mechanics: A Nonequilibrium Ensemble Formalism*, Kluwer Academic Publishers, Dordrecht, 2002) e não nos deteremos aqui nesta questão; é suficiente dizer que a construção deste operador está baseada no critério de Jaynes [7,8] (baseado por sua vez nas idéias de Shannon sobre a teoria matemática da informação) de maximização da entropia estatística de Gibbs [9], isto é, $S(t) = -\text{Tr}\{\hat{\rho}(t)\ln\hat{\rho}(t)\}$, com os vínculos que impõe o conhecimento parcial que se tem sobre o sistema (informação ao nível macroscópico imposta pelas condições num dado experimento), a todo instante $t'$, no intervalo $t_0 < t' < t$, entre o instante inicial de preparação do sistema, $t_0$, e o instante final, $t$, no qual se faz a medida de um dado experimento.

Chamamos a atenção ao importante fato de que o mesmo princípio variacional, quando aplicado a sistemas em equilíbrio com reservatórios, proporciona as bem conhecidas distribuições de probabilidade de Gibbs, neste caso, de equilíbrio. Também comentamos que o operador ou distribuição de não equilíbrio que o formalismo proporciona, quando o sistema é desacoplado de todas as perturbações externas que o afastam do equilíbrio e é deixado somente em contato com um dado conjunto de reservatórios, o operador $\hat{\rho}(t)$ tende assintoticamente no tempo para a correspondente distribuição de Gibbs em equilíbrio.

O Operador Estatístico de Não Equilíbrio $\hat{\rho}(t)$, na versão devida a D. N. Zubarev (a ser consistentemente usado por nós), é dado por [10,11,12]

$$\hat{\rho}_\varepsilon(t) = \exp\left\{\ln\hat{\bar{\rho}}(t,0) - \int_{-\infty}^{t} dt' e^{\varepsilon(t'-t)} \frac{d}{dt'}\ln\hat{\bar{\rho}}(t',t'-t)\right\}, \qquad (4)$$

onde $\hat{\bar{\rho}}$ é um operador auxiliar que tem a forma

$$\hat{\bar{\rho}}(t,0) = \exp\left\{-\phi(t) - \sum_{j=1}^{n} \int d^3r\, F_j(\vec{r},t)P(\vec{r})\right\}, \qquad (5)$$

e

$$\hat{\bar{\rho}}(t',t'-t) = e^{-(t'-t)\hat{H}/i\hbar}\hat{\bar{\rho}}(t',0)e^{(t'-t)\hat{H}/i\hbar}.$$

O parâmetro $\varepsilon$ que aparece na equação (4) é uma quantidade infinitesimal positiva que vai para zero ($\varepsilon \to {}_{+}0$) após os cálculos dos traços serem efetuados. Este processo de limite está relacionado com a quebra de simetria temporal na equação de Liouville, o que permite selecionar só as soluções retardadas, garantindo desta forma, a irreversibilidade das equações de evolução e, portanto, uma correta descrição do processo dissipativo, que é, insistimos, uma hipótese *ad hoc*. Chamamos a atenção ao fato de que este procedimento no Método do Operador Estatístico de Não Equilíbrio introduz irreversibilidade a partir das condições iniciais e, como dito, imposto pelo formalismo. Na Eq. (5), $\phi(t)$ e o conjunto $F_j(\vec{r},t)$ são os multiplicadores de Lagrange que o método de maximização da entropia de Gibbs introduz. Em particular, $\phi(t)$ garante a normalização de $\hat{\rho}(t)$, isto é, $\text{Tr}\{\hat{\rho}(t)\} = 1$, para todo $t$, ou seja

$$\phi(t) = \ln \text{Tr}\left\{\exp\left(-\sum_{j=1}^{n}\int d^3r\, F_j(\vec{r},t)P(\vec{r})\right)\right\}, \tag{6}$$

A partir da Eq. (4) podemos mostrar que o operador estatístico pode ser separado em dois termos

$$\hat{\rho}_\varepsilon(t) = \hat{\bar{\rho}}(t) + \hat{\rho}'(t), \tag{7}$$

onde o operador auxiliar $\hat{\bar{\rho}}(t)$ dá o valor instantâneo no tempo das macrovariáveis e não contém os efeitos de relaxação. Por outro lado, $\hat{\rho}'(t)$ é o responsável pela descrição da evolução irreversível do sistema [13,14] e, portanto, está incluído no cálculo da derivada temporal de $Q_j(\vec{r},t)$. Derivando no tempo a equação (1) obtemos as equações de evolução para as macrovariáveis $Q_j(\vec{r},t)$,

$$\frac{d}{dt}Q_j(\vec{r},t) = \frac{1}{i\hbar}\text{Tr}\{[\hat{P}_j(\vec{r}),\widehat{H}]\}\hat{\rho}_\varepsilon(t), \tag{8}$$

isto é, são as equações mecânico quânticas de Heisenberg para as quantidades $\hat{P}_j(\vec{r})$ promediadas sobre o ensemble de não equilíbrio caracterizado por $\hat{\rho}_\varepsilon(t)$. Usando as Eqs. (2), (3) e (7) mostra-se que

$$\frac{d}{dt}Q_j(\vec{r},t) = J_j^{(0)}(\vec{r},t) + J_j^{(1)}(\vec{r},t) + \Im_j(\vec{r},t), \tag{9}$$

onde

$$J_j^{(0)}(\vec{r},t) = \frac{1}{i\hbar}\text{Tr}\{[\hat{P}_j(\vec{r}),\widehat{H}_0]\}\hat{\bar{\rho}}(t), \tag{10}$$

$$J_j^{(1)}(\vec{r},t) = \frac{1}{i\hbar}\text{Tr}\{[\hat{P}_j(\vec{r}),\widehat{H}_1]\}\hat{\bar{\rho}}(t), \tag{11}$$

$$\Im_j(\vec{r},t) = \frac{1}{i\hbar}\text{Tr}\{[\hat{P}_j(\vec{r}),\hat{H}_1]\}\hat{\rho}'(t), \qquad (12)$$

Os termos $J_j^{(0)}(\vec{r},t)$ e $J_j^{(1)}(\vec{r},t)$ estão associados ao operador auxiliar $\hat{\bar{\rho}}(t)$ que, como dito anteriormente, determina o valor instantâneo das macrovariáveis. O termo $\Im_j(\vec{r},t)$ está relacionado com $\hat{\rho}'(t)$ e é o responsável pela evolução irreversível do sistema [3,13]. Uma teoria perturbativa completa que permite calcular o termo $\Im_j(\vec{r},t)$ na ordem de interação $\hat{H}_1$ desejada foi obtida por Lauck et al. [15,16]. Formalmente este termo pode ser escrito na forma

$$\Im_j(\vec{r},t) = \sum_{n=2}^{\infty} J_j^{(s)}(\vec{r},t), \qquad (13)$$

onde o índice ($s$) indica a ordem das interações presentes em cada termo. Ppensando em termos de um gás de moléculas corresponderia à contribuição resultante da colisão de $s$ partículas em uma teoria cinética tradicional. Aqui vamos apresentar só os resultados do limite Markoviano (interação fraca) que consiste num desenvolvimento até a segunda ordem nas interações, que está dado pelo termo $J_j^{(2)}$, ou seja, $\Im_j(\vec{r},t) \simeq J_j^{(2)}(\vec{r},t)$, com

$$J_j^{(2)}(\vec{r},t) = -\frac{1}{\hbar^2}\lim_{\varepsilon\to{}_+0}\int_{-\infty}^{0}dt'\,e^{\varepsilon t'}Tr\left\{\left(\left[\hat{H}_1(t'),[\hat{H}_1,\hat{P}_j(\vec{r})]\right]+i\hbar\sum_m\hat{P}_m(\vec{r})\frac{\delta J_j^{(1)}(\vec{r},t)}{\delta Q_m(\vec{r},t)}\right]\right)\hat{\bar{\rho}}(t)\right\}, \qquad (14)$$

com $\delta$ indicando a derivada funcional.

Esta aproximação descreve muito satisfatoriamente uma grande parte de vários resultados experimentais na área de sistemas afastados do equilíbrio (certos aspectos no caso de polímeros, semicondutores e sistemas biológicos). Em (14) a dependência temporal explícita em $\hat{H}_1$ é dada por

$$\hat{H}_1(t') = e^{-t'\hat{H}_0/i\hbar}\hat{H}_1 e^{t'\hat{H}_0/i\hbar}. \qquad (15)$$

Assim, conhecendo as equações de transporte para as macrovariáveis de base podemos obter as equações de evolução para os multiplicadores de Lagrange, os quais têm o papel de variáveis termodinâmicas intensivas e descrevem completamente o estado termodinâmico de não equilíbrio do sistema, assim como o fazem as macrovariáveis $Q_j(\vec{r},t)$. Para maiores detalhes sobre a teoria cinética aqui descrita veja: *A Kinetic Theory for Nonlinear Quantum Transport*, C. G. Rodrigues, R. Luzzi, A. R. Vasconcellos, in: Transport Theory and Statistical Physics **29** (7), 733 (2000).

Quanto à viabilidade técnica existem diversas aplicações com muito êxito no estudo de semicondutores utilizando o Método do Operador Estatístico de Não Equilíbrio, como por exemplo, os listados nas referências [17-83].

## Referências